\documentclass[12pt]{article}\usepackage[hyperfootnotes=false]{hyperref}
\usepackage{epsfig}
\usepackage{float}

\usepackage{caption}

\usepackage{amsmath}
\usepackage{amssymb}
\usepackage{graphicx}
\newlength{\abstractwidth}
\renewcommand{\thefootnote}{\fnsymbol{footnote}}
\renewcommand{\thanks}[1]{\footnote{#1}} 
\newcommand{\starttext}{
\setcounter{footnote}{0}
\renewcommand{\thefootnote}{\arabic{footnote}}}
\renewcommand{\theequation}{\thesection.\arabic{equation}}
\newcommand{\be}{\begin{equation}}
\newcommand{\bea}{\begin{eqnarray}}
\newcommand{\eea}{\end{eqnarray}}
\newcommand{\beq}{\begin{equation}}
\newcommand{\ee}{\end{equation}}
\newcommand{\eeq}{\end{equation}}

\def\ba{\begin{eqnarray}}
\def\ea{\end{eqnarray}}

\def\12{{1 \over 2}}

\def\la{\langle}
\def\ra{\rangle}

\def\simleq{\; \raise0.3ex\hbox{$<$\kern-0.75em
\raise-1.1ex\hbox{$\sim$}}\; }
\def\simgeq{\; \raise0.3ex\hbox{$>$\kern-0.75em
\raise-1.1ex\hbox{$\sim$}}\; }

\def\O2{\Omega_2}

\def\bi{\begin{itemize}}
\def\ei{\end{itemize}}

\def\sc{\setcounter{equation}{0}}

\def\W{$\Omega$}
\def\W'{$\Omega$}

\def\V{\Omega}
\def\V'{\Omega}

\def\O{${\cal{O}}$}

\def\c{{\cal{C}}}

\def\bn{\bigskip \noindent}

    \def\cg{$\c$-geometry}
     \def\cg2{$\c_2$-geometry}

\makeatletter
\g@addto@macro\normalsize{%
  \setlength\abovedisplayskip{10pt}
  \setlength\belowdisplayskip{20pt}
  \setlength\abovedisplayshortskip{10pt}
  \setlength\belowdisplayshortskip{20pt}
}
\makeatother

\usepackage{color}

\begin{document}
\renewcommand{\theequation}{\thesection.\arabic{equation}}
\begin{titlepage}
\rightline{}
\bigskip
\bigskip\bigskip\bigskip\bigskip
\bigskip
\centerline{\Large \bf {Black Holes and Complexity Classes}}

\bn

\bigskip
\begin{center}
\bf   Leonard Susskind  \rm

\bigskip

 Stanford Institute for Theoretical Physics and Department of Physics, \\
Stanford University,
Stanford, CA 94305-4060, USA \\
\bigskip

\end{center}

\begin{abstract}

It is not known what the limitations are on using quantum computation to speed up classical computation. An example would be the power to speed up PSPACE-complete computations.  It is also not known what the limitations are on the duration of time over which classical general relativity can describe the interior geometry of black holes. What is known is that these two questions are closely connected: the longer GR can describe black holes, the more limited are quantum computers. This conclusion, formulated as a theorem, is a result of unpublished work done by Scott Aaronson and myself which I explain here.

\medskip
\noindent
\end{abstract}


\end{titlepage}

\starttext \baselineskip=17.63pt \setcounter{footnote}{0}

\section{Introduction}
Several years ago I askd Scott Aaronson  a question: Can it be proved that the complexity of a universal quantum circuit, such as those that have been conjectured to describe black holes, grows at the fastest possible rate---linearly with time---until it saturates at the maximum complexity (exponential in the number of qubits)? The result was a theorem (mainly due to Aaronson) connecting the growth of complexity with certain plausible properties of complexity classes. In this note I explain the theorem and its importance in a way that I hope can be understood both by physicists and complexity theorists.

I will begin by explaining intuitively why one might expect such a connection. Let us suppose that a universal quantum circuit, when run for exponential time failed to produce complexity greater than polynomial in the number of qubits. Consider some problem which is classically hard, i.e., it takes exponential time $\sim c^N$ to solve it. We could obviously solve the problem in time $c^N$ with the quantum computer by running it in classical mode. However, by assumption the (quantum) complexity at time $c^N$ is polynomial in $N$. It follows that there is a way to get to the answer in a polynomial number of steps running the computer as a quantum computer.

As we will see the argument also goes in the other direction: If the universal  quantum computer produces greater than polynomial complexity in exponential time, then certain hard problems (PSPACE-complete) cannot be solved in polynomial time by a quantum computer.
Later this will be formulated as a precise theorem.

What does this have to do with black holes? We will see shortly.

\sc
\section{Two Conjectures}

\begin{figure}[H]
\begin{center}
\includegraphics[scale=.3]{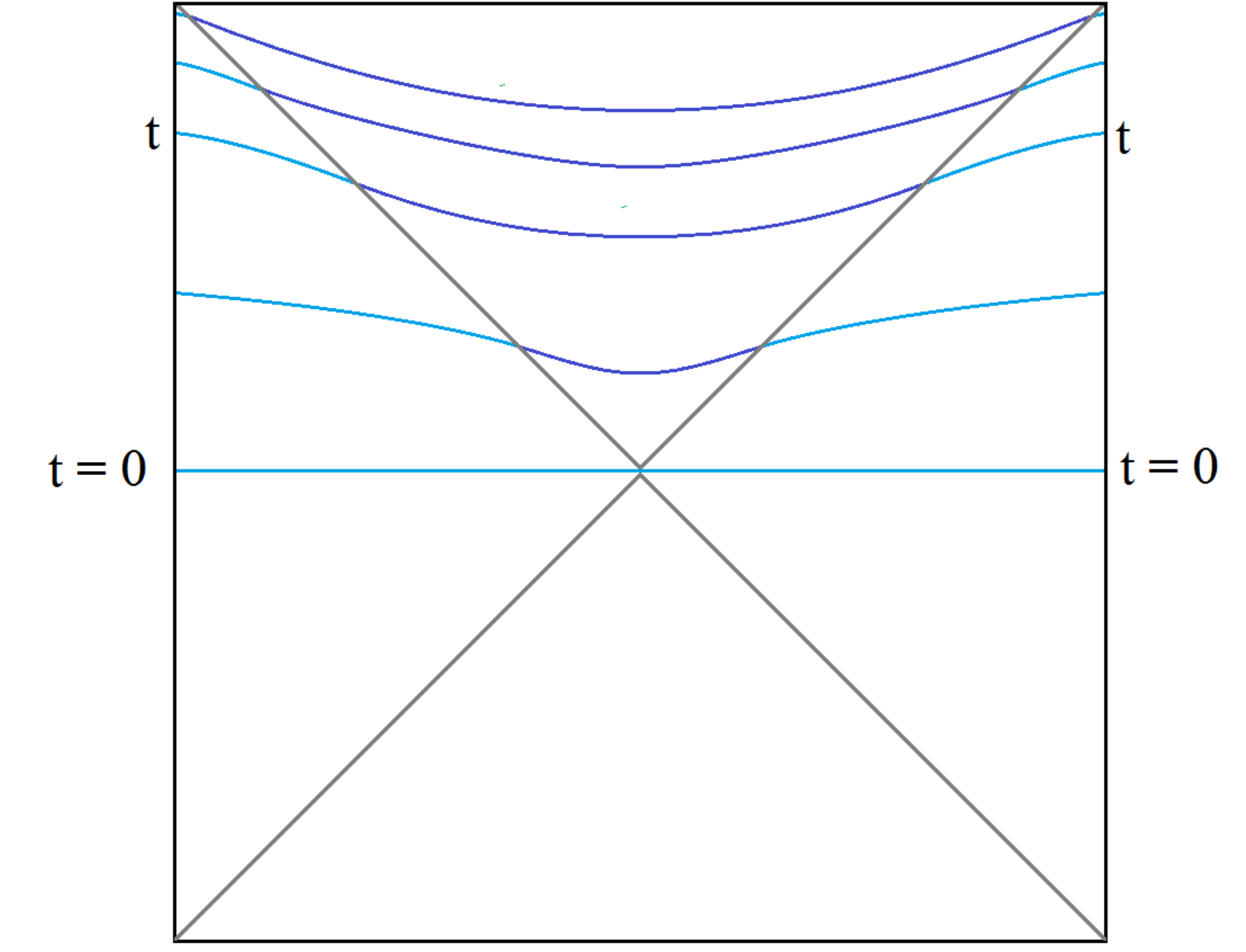}
\caption{Penrose diagram for a ``two-sided" black hole. The spacetime is sliced by maximal space-like slices anchored on the boundaries. It is obvious that the portions of the slices behind the horizon grow with time. }
\label{f1}
\end{center}
\end{figure}

According to  classical general  relativity  the volume of space behind the horizon of a black hole   grows  linearly (with time), into the eternal future \cite{Susskind:2014rva}. The two-sided   black hole    in gauge-gravity duality (AdS/CFT)  \cite{Maldacena:2001kr} is the best studied example of this growth, and we will refer to it throughout. The Einstein-Rosen bridge (ERB)  connecting the black holes on either side grows so that its length, volume, and action, all increase proportional to the time defined on the AdS boundaries. The phenomenon is illustrated in the Penrose diagram shown in blue in figure \ref{f1}. The spacetime is sliced by smooth space-like slices\footnote{A useful geometric way of defining the slices is to make them maximal. That means for given anchoring time at the boundaries the space-like slices should have maximum spatial volume. An alternative quantity is the action of a Wheeler-DeWitt patch as explained in \cite{Brown:2015bva}\cite{Brown:2015lvg}.} anchored at the boundary at a series of increasing times.
At a particular time $t$ the ERB is defined as the intersection of the slice $t$ with the interior region. In other words the ERB is the portion of the slice (shown in darker blue in figure \ref{f1}) that lies behind the horizon. It is evident from the figure that the 
ERB grows with time and an easy general relativity calculation \cite{Susskind:2014rva} shows that the growth is linear in time.

When quantum theory is introduced it becomes natural to ask what property of the instantaneous Schrodinger-picture state  encodes the size of the ERB? The  proposed answer is quantum computational complexity \cite{Susskind:2014rva}\cite{Stanford:2014jda}\cite{Roberts:2014isa}\cite{Susskind:2014moa}\cite{Brown:2015bva}\cite{Brown:2015lvg}, also known as gate complexity or circuit complexity\footnote{From now on I will just use the term complexity.}. By definition it is
the minimum number of gates needed to prepare the quantum state from some initial simple state.

The indefinite growth of an ERB predicted by general relativity  is an artifact of classical physics. On a sufficiently large time scale a closed system of finite entropy must undergo quantum recurrences. Whatever the connection is between the quantum state and the size of the ERB, on quantum recurrence time scales it must be quasiperiodic---a distinctly non-classical behavior. If the connection is through complexity, the time scale for  general relativity to break down has to be less than some exponential of the entropy $S,$  since the maximum possible complexity for a closed system is exponential in $S$. 

In this paper we will consider the relation between two kinds of conjectures: on the one hand, conjectures about the time scale for the breakdown of classical gravity; and on the other hand, conjectures about  the inclusion properties of certain quantum and classical complexity classes. The basic gravitational conjecture can be stated as follows:

\bn
\bf Conjecture: \rm 

\noindent
\it
Classical general relativity governs the behavior of an ERB for as long as possible.
\rm

\bn
This means that the size of the ERB grows linearly  for as long as quantum mechanics allows. Assuming the duality between ERB geometry and complexity,   the linear growth continues until the volume is exponential in the entropy $S.$ Stated in terms of complexity, the complexity of the quantum state grows linearly until it reaches its maximum possible value  $e^{S}$

If, as believed,  black holes can be modeled as systems of qubits evolving by unitary evolution, the conjecture  can be re-stated in the form:

\bn
\bf Conjecture: \rm 

\noindent
\it The complexity of certain quantum circuits grows linearly with the number of time-steps, until it reaches its maximum value,  exponential in the number of qubits.
\rm

\bn

Here, the phrase, certain quantum circuits, refers to the type of circuits that govern the evolution of a black hole in anti de Sitter space. In particular we expect that they are computationally universal.

The conjecture has interesting implications for complexity theory. I  will state one here and 
generalize it later. It implies that the classical complexity class PSPACE is not contained in the quantum complexity class  BQP/poly.  
Roughly stated, there are problems that can be solved by classical circuits with width $N$ and arbitrary depth that cannot be solved with a quantum computer in polynomial time with polynomial advice\footnote{See appendix for definition of advice.}. 
 Thus a physics conjecture about the limits of classical gravity is directly related to a conjecture about the limits of quantum computation:

\bn
\bf Conjecture \rm 

\noindent
\it
PSPACE  is not contained in BQP/poly.
\rm

\bn

\bn

There are more  general statements that tend even more strongly in the same direction, namely the longer the linear growth of ERBs prevails, the less powerful quantum computers are for solving PSPACE complete problems.

\sc
\section{Qubits and Black Holes}
Here we will review a few relevant things about the qubit description of black holes. It is commonly believed that black holes can be modeled as systems of $N$ qubits with $N$ is of order the entropy of the black hole. (An example is the SYK model \cite{Sachdev:1992fk}\cite{Kitaev} at infinite temperature.)
The initial state of the two-sided system is the maximally entangled state of $2N$ qubits which can be written as a product of $N$ Bell pairs shared between the left and right sides. This is the  thermofield-double state at infinite temperature. It may also be written in the form,
\be 
|\Psi(0) \ra = \sum_{ij} \delta_{ij}|i, j\ra
\ee
where $i,j$ label basis states in the left and right Hilbert spaces.

The system evolves with time so that after time $t$ the state becomes,
\be 
|\Psi(t) \ra = \sum_{ij} U_{ij}|i, j\ra
\ee
where the unitary matrix $U$ is given by,
\be 
U_{ij} = \la i|e^{-2iHt} |j\ra.
\ee

The restricted complexity of the state $|\Psi(t) \ra$ will be defined as the minimal number of 2-local gates that it takes to prepare $|\Psi(t)\ra$ starting with the simple state $|\Psi(0) \ra,$ assuming the following restriction: No gates are allowed to couple the left qubits to the right qubits. This restriction makes sense if we think of the two subsets of qubits as being spatially very far from one another.

With this restriction it is clear that the complexity of the state $|\Psi(t)\ra$ is the complexity of the time development operator $U(2t) = e^{-2iHt}$ for a one-sided system. In other words the restricted state-complexity of the two-sided system is the same as the operator complexity of the one-sided evolution operator. From now on we concentrate of the complexity of $U.$

\bn

 There are many $N$-qubit circuits that can generate $U.$ This is obvious because we can always gratuitously insert a gate and its inverse. Therefore  the number of gates in the circuit that generates a given unitary operator $U$  is not a well defined concept. What is well defined is the number of gates in the \it smallest  \rm quantum circuit that generates $U.$ The smallest circuit is called $A_U$ and the number of gates in it is the complexity of $U.$ Call it 
  $$\c_U.$$

Assume $U $ is generated by some dynamics that can be represented by a quantum circuit that successively repeats a low-depth circuit $u$ in the form  $u^t$ ( $u$ raised to the power $t$). The number of gates in $u^t$ is an upper bound on $\c (U)$ and obviously grows linearly with $t$.

Our goal is to understand the complexity of $|\Psi(t)\ra$ which as we've seen is the same as the complexity of $U(2t)$
\be
\c(|\Psi(t)\ra)= \c_{U}(2t).
\label{3.1}
\ee


The black hole conjecture says that the complexity of $U(t)$ grows linearly with $t$ for a time exponential in $N.$


\sc
\section{The Complexity Hypotheses}

We will be interested in the complexity of $U$ after an exponential time $t=c^N$   with $1<c<2,$ 

$$\c_U(c^N).$$

 By the \it weak complexity hypothesis \rm (WCH) we will mean that $\c_U(c^N)$  grows faster than any power of $N.$


\bn

\bf  {\large{Theorem 1}} : \rm

\bn
The WCH is true if and only if PSPACE is not contained in BQP/poly.
Equivalently we can state this in contrapositive form:

\bn
PSPACE is not contained in BQP/poly iff  the  WCH is false.

\sc
\section{Proof of Theorem 1}

\subsection{The L Problem}
In proving Theorem 1 it will be useful to define a PSPACE-complete problem. Proving   a PSPACE-complete problem is in a given complexity class (BQP/poly for example)    proves that all of PSPACE is in that class. 
Thus we begin by defining a specific PSPACE-complete problem $L:$

\begin{figure}[H]
\begin{center}
\includegraphics[scale=.4]{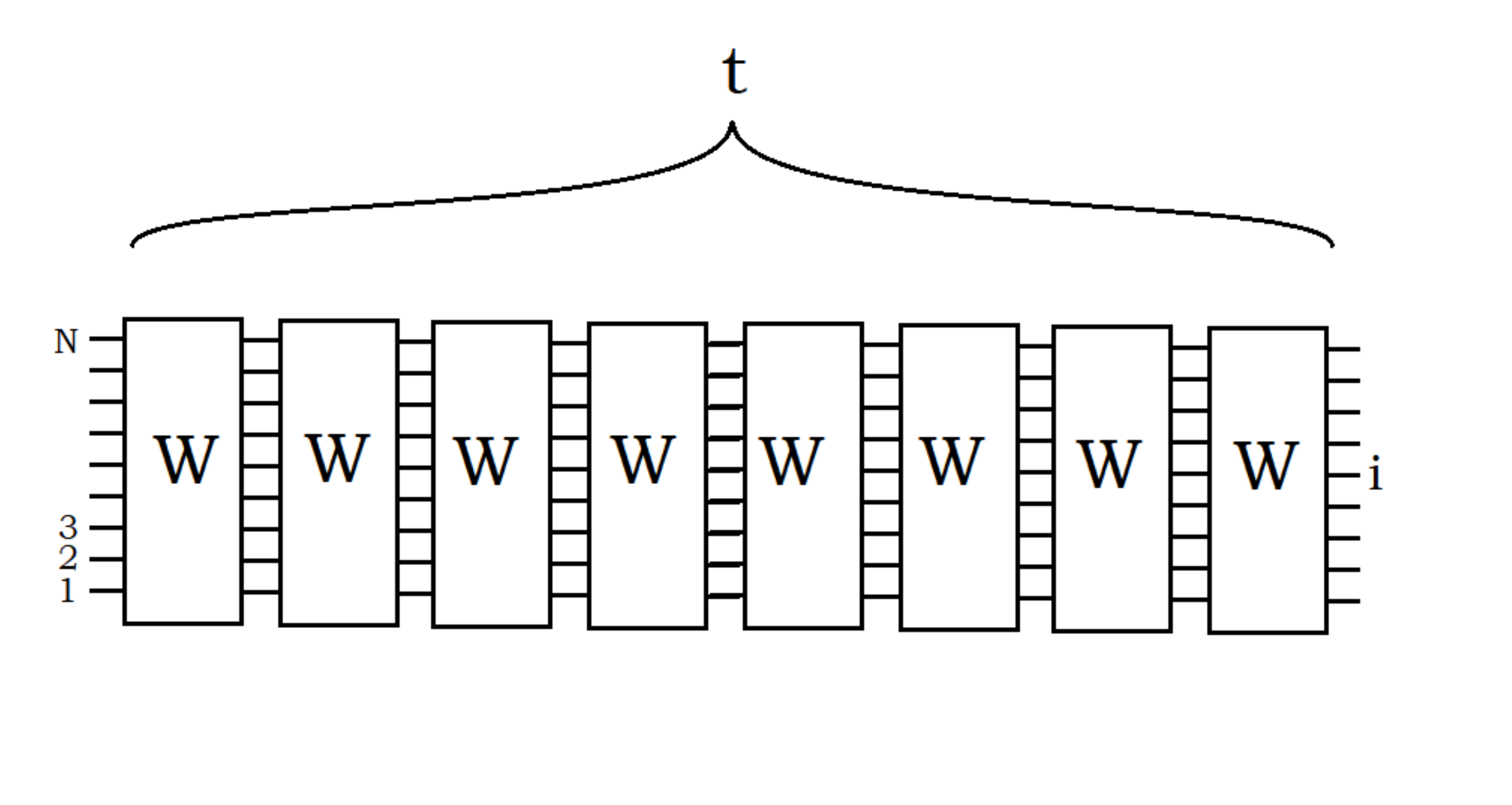}
\caption{A circuit built by repeating a finite depth circuit $W$. $W^t$ is  the unitary evolution  operator after $t$ discrete time steps.}
\label{WWWW}
\end{center}
\end{figure}

Let $W$ be a unitary transformation on $N$ qubits, which applies a single step of a reversible computationally-universal classical cellular automaton (one for which predicting the behavior of $N$ cells after $\sim 2^N$ time steps is a PSPACE-complete problem).  There are many examples of such automata (and the assumption of reversibility doesn't harm the PSPACE-completeness, by a theorem of Lange, McKenzie, and Tapp   \cite{Lange}).
Let $W^t$ be the result of applying $W$ repeatedly  $t$ times (see figure \ref{WWWW}. 

Let the input state for the cellular automaton 
 be a bit-string $x$ of length $N.$ Sometimes we will call the bit-string $|x\ra $ when we want to emphasize its role as a quantum state.   The problem $L$ is to determine the $ith$ bit of the string

$$(W)^{c^N}|x\ra.$$

\bn
 This problem is PSPACE-complete.

In order to label things precisely we will use the index $i = 1,.....,N$  to label the bits.  The problem  $L_i$ is that of determining the ith bit of $$W^{c^N}|x\ra.$$
First suppose that PSPACE is contained in BQP/poly  (quantum computers are powerful). Under that hypothesis we will show that there is a polynomial-size quantum circuit $A$ that can implement $W^{c^N}.$

By hypothesis the problem $L_i$ 
can be solved by a polynomial-size unitary quantum circuit which we will call $C_i.$ What the circuit $C_i$ does when it act on $|x\ra$ is to produce a state in which the first slot of the register contains the  ith  bit of $W^{c^N}|x\ra.$  What occurs in the other slots is a definite quantum state $|\psi_i\ra$ but we will not need its form.

It is of course necessary to encode the protocol for $C_i$ as additional data. This information is the \it advice \rm indicated by the notation /poly. Since by assumption $C_i$ is polynomial in $N$ and the index $i$ runs over $N$ values, the advice is indeed polynomial.

Furthermore the hypothesis implies that there is a second polynomial quantum circuit $B_j$ (also part of the advice) which can be applied to $A|x\ra $ which will give the jth bit of $|x\ra .$ We want to show that there is a polynomial quantum circuit that takes $|x\ra $ to $W^{c^N}|x\ra$  for any input $x.$ In other words there is a polynomial circuit that computes  $W^{c^N}.$

The construction is as follows:

First copy the string $|x\ra$ $N$ times  into registers $r_1, r_2, ...., r_N.$ Note that this is not forbidden quantum-cloning since the string $x$ is given in the classical or computational basis\footnote{ In order to label the copies we  from $1$ to $N$ we also need to add some additional bits. The additional number is small and can be ignored.}

Now act in parallel with  $C_i$ on the ith copy of $|x\ra.$  The result will be that the $ith$ register will have as its first entry the $ith$ bit of $W^{c^N}|x\ra.$  Now it is a simple matter to sequentially copy the the first bit of each register into a special output register $r_o,$ thus encoding $W^{c^N}|x\ra$ in the register $r_0.$

However we are not finished. We need to get rid of all the left over junk in $r_1, r_2, ...., r_N.$

Each of these registers can be acted upon by $C^{-1}$ to return it to $|x, i \ra.$ Since $|x\ra$ is diagonal in the computational basis, the many copies can be classically ``condensed" to a single copy.

\begin{figure}[H]
\begin{center}
\includegraphics[scale=.7]{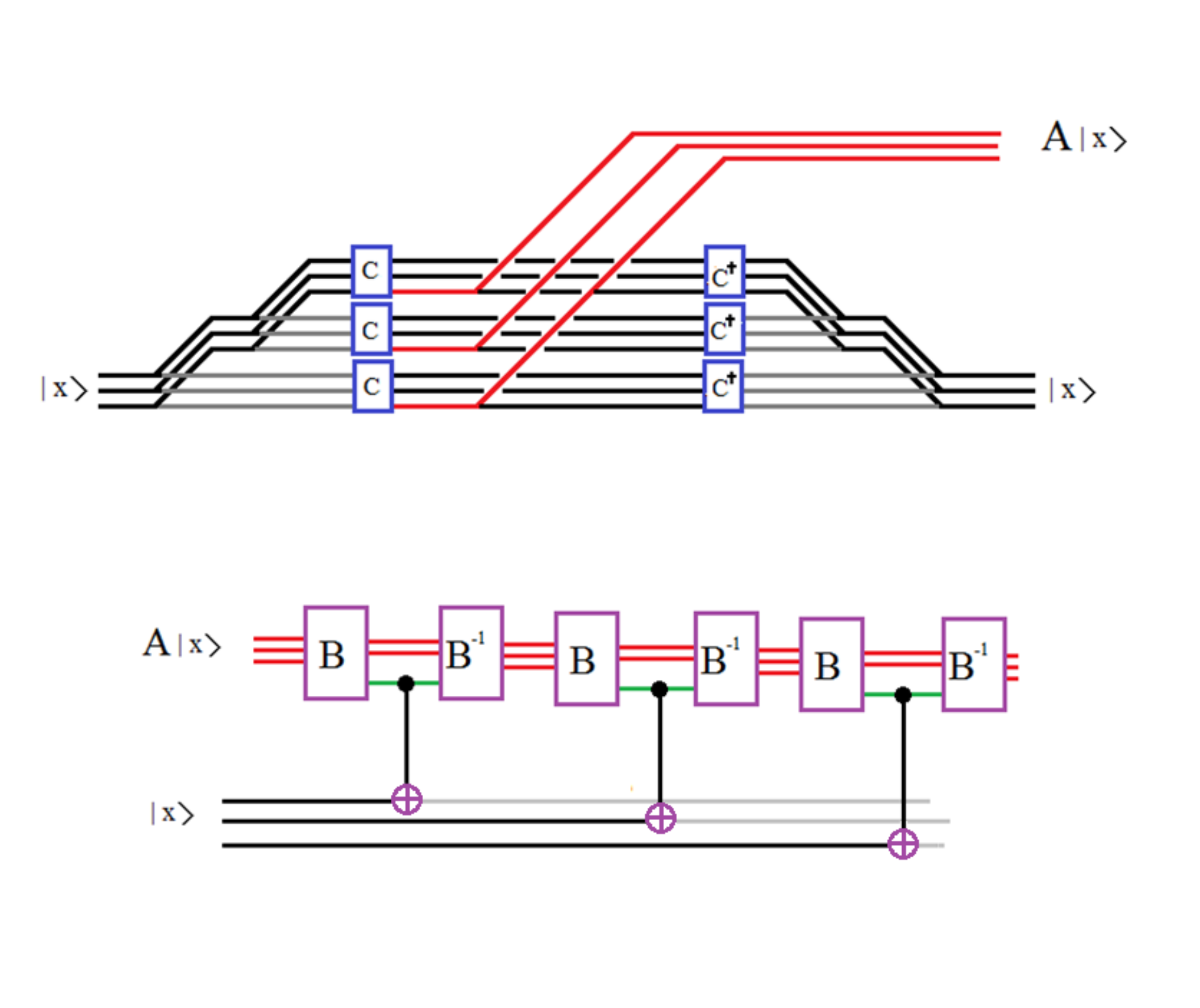}
\caption{Polynomial size circuit described in the text that can compute $A|x\ra$ under the stated hypothesis that $L_i$ can be solved by a polynomial size quantum circuit $C.$}
\label{f3}
\end{center}
\end{figure}

The steps are shown in the first circuit of figure \ref{f3}.

Finally the last part of the protocol is to transform each of the bits of $|x\ra $ to $0$ thus effectively erasing them.  This can be done by the addition of the cicuit shown in the lower half of figure \ref{f3}.
The resulting circuit acts on any $|x\ra$ to give $W^{c^N}|x\ra.$ Moreover the total  number of gates is polynomial in $N.$

Thus we have shown that under the stated hypothesis, the complexity of $W^{c^N}$  is polynomial. This proves that  PSPACE-in-BQP/poly   implies that the complexity of $W^{2N}$ is polynomial. Equivalently, the hypothesis that the complexity of $W^{2N}$ is greater than any polynomial implies that PSPACE is not in BQP/poly.

\bn








Next we need to prove that the hypothesis $\c_{W^{2^N}} $ in polynomial implies that  PSPACE is  contained in BQP/poly. This is straightforward. Suppose there is a polynomial quantum circuit that implements $W^{2N}.$ Then it can be used to solve the PSPACE-COMPLETE problem $L$ in an obvious way.  Under the stated hypothesis this proves that PSPACE is in BQP/poly.

Thus we have proved Theorem 1: The  weak complexity hypothesis is equivalent to  the statement that  PSPACE is not contained in BQP/poly. 

\sc
\section{Stronger Hypothesis}

That PSPACE should not be in BQP/poly is not a surprise and is considered highly plausible. But if we assume the conjecture that complexity grows linearly with time until it reaches its maximum value, there are much stronger implications. For example suppose we replace the WCH with the stronger assumption that $C_U(c^N)$ grows faster than any sub-exponential. Call it the stronger complexity hypothesis (SerCH). Then by the same reasoning as in the previous section we prove:

\bn

\bf  {\large{Theorem 2}} : \rm

\bn
The SerCH is true if and only if PSPACE is not contained in BQSUBEXP/subexp.
Equivalently we can state this in contrapositive form:

\bn
PSPACE is not contained in BQSUBEXP/subexp iff  the  SerCH is false.

\bn
Although PSPACE not in BQSUBEXP/subexp may be plausible it is a much stronger statement than PSPACE not in BQP/poly. For that reason complexity theory cannot be said to confirm the black hole conjecture. On the other hand, the assumption that general relativity holds for a long as possible would put very strong constraints on complexity theory and implies PSPACE not in BQSUBEXP/subexp.

\section{Conclusion}

We have not proved that the complexity of a universal quantum circuit becomes maximally large after an exponential time. This is a very difficult problem. From Nielsen's geometric approach to complexity \cite{Nielsen} it is possible to prove that $\c(t)$ increases linearly for some non-zero time interval, but that’s all. We have been able to make a small step by relating the issue to a conjecture about complexity classes.

The problem was motivated by a  question about  
the limits of classical general relativity: For how long a time does classical GR hold during the evolution of a black hole?  This connection between black holes and complexity classes is unexpected, and in my opinion very remarkable. Broadly speaking it says that the longer classical general relativity describes the interior of black holes, the less quantum computers have power to solve PSPACE-complete problems.

\section*{Acknowledgements}

The author thanks Scott Aaronson for collaboration and patient mentoring.  Support for this research came through NSF Award Number 1316699.

\appendix

\section{Some Black Hole Terminology}

In this appendix a brief explanation of some terminology is given, first black holes:

\bi

\item AdS: Anti de Sitter space is a spacetime with uniform negative curvature. It may contain black holes. AdS has a causal boundary. 

\item AdS/CFT: Quantum gravity in AdS is described by a conformal quantum field theory (CFT)  located on the boundary. The duality between quantum gravity in AdS and quantum field theory on the boundary is often called gauge/gravity duality. A black hole in AdS is dual to a thermal state of the CFT.

\item Einstein-Rosen bridge: An ERB is a wormhole connecting two entangled black holes at either of the ends of the wormhole. The ERB grows in length with time. Classically it grows forever.

\item Thermofield Double state: The quantum state of a pair of entangled black holes at time $t=0$ is called the Thermofield Double (TFD). As time evolves to greater complexity corresponding to the growth of the volume of the ERB. In gauge/gravity duality the TFD is a state of a pair of CFTs thought of as living on two disconnected and non-interacting boundaries. The only connection between the two CFTs is that they are entangled.

\ei

\section{Some Complexity Terminology}
I will now give a brief explanation of some complexity terminology which occurs in the main body of this paper.
More complete information can be found in ``The Complexity Zoo"  \cite{zoo}

\bi

\item Input Size: The input data for a problem is usually in the form of a bit-string. The size on an instance $n$ is the number of bits in the input string for that instance.

\item Decision problems: Problems with a yes-no answer. Example: Is the $i^{th}$ decimal coefficient of $\pi$ even? Note that the term decision problem does not apply to a single instance but to an infinite set of problems.

\item Advice: Advice means a set of bit-strings which serve as information that a computer can refer to. Generally there is not an advice bit-string for each instance of a problem. Rather there is a single advice string for each size of the problem. Polynomial advice refers to advice strings that grow no faster than a power of the input size $n.$

\item Non-uniform Advice: Non-uniform advice allows the advice string for each $n$ to be specified independently, i.e., with no uniform rule.

\item PSPACE: The set of decision  problems that can be solved by a Turing machine using a polynomial size memory. Note that the limit is on the memory size, not on the length of time the machine can run.

 \item PSPACE-complete is the class of problems in PSPACE that
 every other problem in PSPACE can be transformed into in polynomial time. If a PSPACE-complete problem is an a complexity class, then PSPACE itself is in that class.
 
  \item BQP:   The class of decision problems solvable by a quantum computer in polynomial time, with an error probability of at most $\epsilon$ for all instances. The number $\epsilon$ is arbitrary. 
  
   \item BQP/poly:   The class of decision problems solvable by a BQP machine supplemented with non-uniform polynomial advice.

   \item  Computational Basis: The class of quantum states that are eigenstates of all qubit $Z$ operators.  They are essentially the same as the states of a classical bit system.
   
   \item Computationally Universal: This refers to the ability of a machine to solve any problem that a Turing machine can solve. More exactly a computationally universal machine can calculate any Turing-computable function.

\ei


 


 
 




\end{document}